\documentclass[12pt,preprint]{emulateapj}
\usepackage{epsfig}
\usepackage{natbib}

\def\deg{$^{\circ}$}

\begin{document}
\title{~~\\ ~~\\ Monitoring the Bi-Directional Relativistic Jets of the Radio Galaxy 3C\,338}
\shorttitle{Monitoring 3C\,338}
\author{ G. Gentile \altaffilmark{1}, C. Rodr{\'{\i}}guez \altaffilmark{1,2}, G. B. Taylor\altaffilmark{1,3}, G. Giovannini \altaffilmark{4,5}, S. W. Allen \altaffilmark{6}, W. M. Lane \altaffilmark{7}, N. E. Kassim \altaffilmark{7}    }

\altaffiltext{1}{Department of Physics and Astronomy, University of New Mexico, Albuquerque, NM 87131, USA}
\altaffiltext{2}{Department of Physics, Universidad Simon Bolivar, Sartenejas, Venezuela}
\altaffiltext{3}{National Radio Astronomy Observatory, Socorro NM 87801, USA}
\altaffiltext{4}{Dipartimento di Astronomia, Universita' di Bologna, 40127 Bologna, Italy}
\altaffiltext{5}{Istituto di Radioastronomia - INAF, 40129
Bologna, Italy}
\altaffiltext{6}{Kavli Institute for Particle Astrophysics and Cosmology, Stanford University, Stanford, CA 94305-4060, USA}
\altaffiltext{7}{Naval Research Lab, Code 7213, Washington, DC, 20375, USA}


\begin{abstract}

We present the analysis of VLA and VLBA observations of the radio
source 3C\,338, associated with the cD galaxy NGC 6166, the central
dominant galaxy of the cluster Abell 2199.  The VLBA observations were
done at 8.4 and 15.4 GHz, while the VLA observations were performed at
0.074, 0.330, and 8.4 GHz.  The milliarcsecond resolution VLBA data,
spanning 7 years, reveal the parsec-scale jets, whose kinematics and
orientation cannot be unambiguously derived.  Based on the observed
morphology, jet/counter-jet length ratio, flux density ratio, and
proper motions of the jet components, we consider two possible
explanations: either the jets are strongly relativistic and lie within
$10^{\circ} - 20^{\circ}$ of the plane of the sky, or they are only
mildly relativistic, and are pointing at an angle between $30^{\circ}
- 50^{\circ}$ from the plane of the sky.  The arcsecond resolution VLA
data enable us to investigate the large scale structure of the radio
source. The morphology of the low frequency radio lobes clearly
indicates that they are associated with the cavities present in the
X-ray emission.  Low frequency observations also reveal an extension
to the south corresponding to an X-ray hole.  The age of these
bubbles, computed from the sound speed, the buoyancy time and the
radiative age are all in fair agreement with each other.  Estimates of
the power necessary to inflate these cavities suggest that the
accretion power onto the central engine has not been constant over
time.

\end{abstract}



\section{Introduction}

The cD galaxy NGC 6166 is the central dominant galaxy in the nearby
($z=0.0304$) cluster Abell 2199.  This galaxy hosts the relatively
powerful radio source, 3C\,338, which emits a total power at
330 MHz of 4.4$\times$10$^{25}$ W Hz$^{-1}$.  This radio source has
been known for a few decades to have an unusual structure on both
large and small scales (e.g. Feretti et al. 1993, Giovannini et
al. 1998).  

On parsec scales 3C\,338 has a compact radio core with two short
($\sim10$ pc), symmetric jets.  Such highly symmetric jets are rare
among 3C (Third Cambridge catalog - Bennett 1962) radio galaxies,
though the few that are known (Hydra A - Taylor 1996; PKS\,2322$-$12 -
Taylor et al. 1999) are also at the centers of large clusters,
suggesting perhaps that the jets in such sources are decelerated on
scales of a few parsecs.  Here we present Very Long Baseline Array
(VLBA) \footnote {The Very Long Baseline Array (VLBA) and the Very
Large Array (VLA) are operated by the National Radio Astronomy
Observatory (NRAO), which is a facility of the National Science
Foundation operated under cooperative agreement by Associated
Universities, Inc.}  observations from six epochs of monitoring
3C\,338 at 8.4 GHz.  Three epochs have also been obtained at 15 GHz.
These observations are used to study the motions of the inner
bidirectional jets in 3C\,338.

On kiloparsec scales 3C\,338 has two symmetric extended radio lobes,
characterized by a steep spectrum ($\alpha$ $\sim$ $-$1.7), and
misaligned with the central emission. The two radio lobes are
connected by a bright filamentary structure.  Polarimetric
observations by Ge \& Owen (1994) revealed strong rotation measure
gradients across most of the extended emission and inferred the
presence of cluster magnetic fields. Both the steep radio spectrum and
strong filamentary emission may be the result of interactions with the
dense intracluster medium.  Further evidence of interaction has been
reported in X-rays where {\it ROSAT} and {\it Chandra} data (Owen \&
Eilek 1998 and Johnstone et al.  2002 respectively) have shown that
the core of the cluster is far from being simple and spherically
symmetric. Johnstone et al. (2002) found evidence for temperature and
metallicity gradients and for structures in the X-ray gas associated
with the radio source. Using new {\it Chandra} data, Sanders \& Fabian
(2006) found an approximately circular X-ray surface brightness
discontinuity at $\sim$ 55 kpc from the center of the cluster, which
they interpret as a weak isothermal shock that might be an important
energy transport mechanism.  Here we present new 74 and 330 MHz VLA
observations, highlighting the existence of
large-scale, diffuse emission connected to the X-ray emitting gas. 


Throughout this discussion, we assume H$_{0}$=70 km s$^{-1}$
Mpc$^{-1}$, $\Omega_m$ = 0.3, $\Omega_\Lambda$ = 0.7, and 1 mas = 0.6 pc.


\begin{figure*}
\centering
\scalebox{0.55}{\includegraphics[angle=-90]{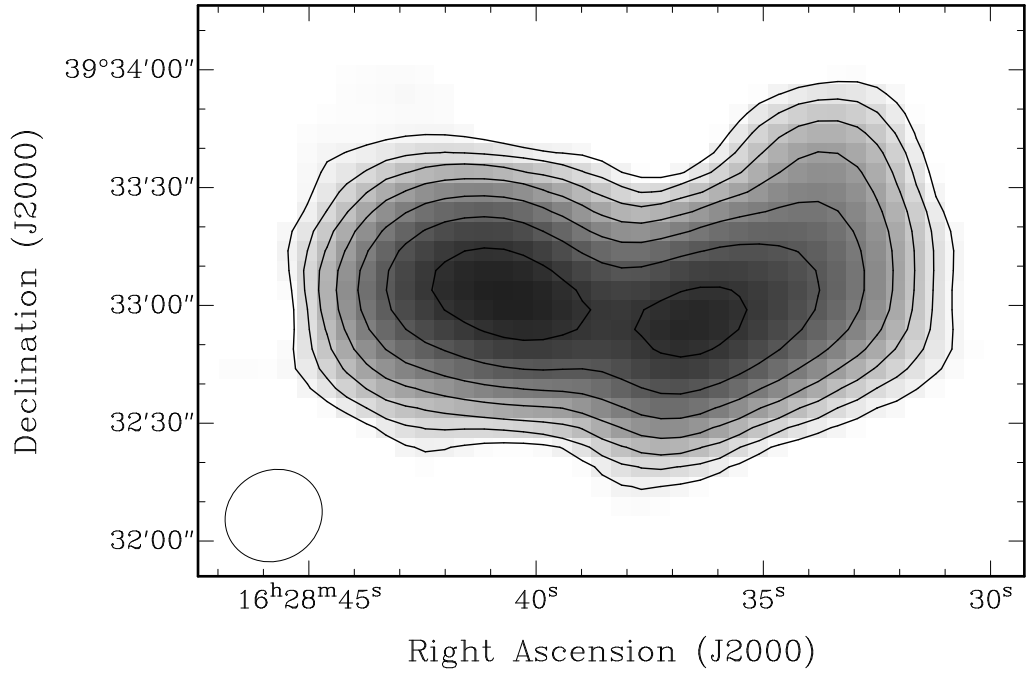}}
\caption{
Contours of the 74 MHz map of 3C\,338. The synthesised beam is shown in the bottom left corner;
its FWHM size is 25.3\arcsec$\times$22.9\arcsec in PA -60$^{\circ}$.
Contour levels are 0.35 (5$\sigma$), 0.7, 1.4, 2.8, 5.6, ... Jy beam$^{-1}$.
The peak flux is 32.7 Jy/beam.
}
\label{74}
\end{figure*}

\begin{figure*}
\centering
\scalebox{0.6}{\includegraphics[angle=-90]{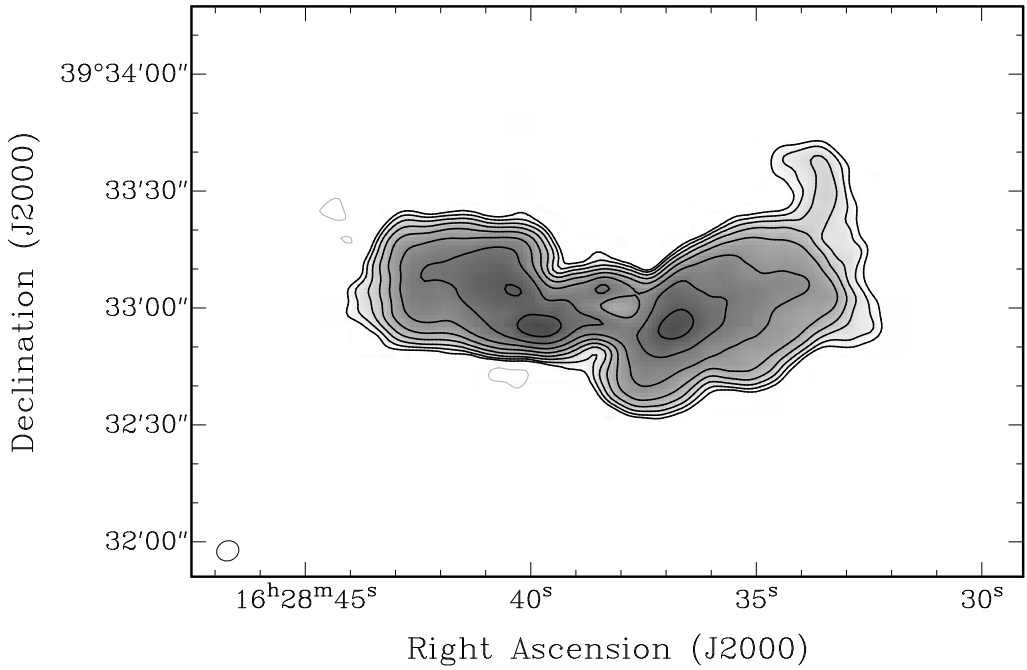}}
\caption{
Contours of the 330 MHz map of 3C\,338. The synthesised beam is shown in the bottom left corner;
its FWHM size is 5.8\arcsec$\times$5.1\arcsec in PA -67$^{\circ}$.
Contour levels are -5 (grey contours), 5 (4.5$\sigma$), 10, 20, 40, 80,  ... mJy beam$^{-1}$.
The peak flux is 0.9 Jy/beam.
}
\label{330}
\end{figure*}

\section{Observations and Data Reduction}
\label{observations}

The 74 and 330 MHz observations were performed on April 16, 2002
at the VLA in the A configuration. At both frequencies the source was observed in spectral line mode 
in order to permit removal of narrow-band radio frequency interference (RFI), 
which at those frequencies
severely affects the data, as well as to mitigate the effects of bandwidth smearing. 
Primary calibration and bandpass corrections for the 330 MHz and 74
MHz data were made using short observations of 3C286 and 3C405
respectively.  The data were edited for radio frequency interference
and averaged in frequency to reduce the number of channels.  Phases
from several observations of 3C286 were transferred to the 330 MHz
data to provide an initial calibration, while the final 330 MHz image
was used as a model to estimate inital phase calibration for the 74
MHz data.  Both datasets were further processed using several
iterative loops of imaging and phase self-calibration followed by two
loops of imaging and amplitude self-calibration with the standard AIPS
tasks IMAGR and CALIB.  Although 3C338 dominates its field, imaging
was done using multiple facets in a wide-field mode at both
frequencies for increased model accuracy during the final amplitude
self-calibrations.

At low radio frequencies, the ionosphere can introduce a refractive
phase error that causes a gross position offset in an image.  This
cannot be removed via the self-calibration process because the phase
error is common to all antennas.  As a result, small angular offsets
are typical in low-frequency maps processed by this method.  When
absolute coordinate accuracy is important it is standard practice to
remove the offset by registering point sources in the image to known
coordinates.  For these data we removed a roughly 10" shift by
comparing the field sources to the NVSS catalog.

Data reduction was also performed using AIPS for the 8.4 GHz VLA observations
in the A, B, CnB, and C configurations.
The data were taken
over the course of five epochs: 
1994 Nov. 17, 1995 Sept. 11, 1998 Dec 02, 2000 Feb. 26, and 2001 Jun 01 (See Table \ref{VLA}).
Standard calibration procedures were performed, as well as imaging and self-calibration. Once
all the data were properly calibrated and the final images obtained for each epoch were 
satisfactory, 
all the visibilities from the five epochs were combined using the task DBCON.

\begin{deluxetable*}{lccccccc}
\tabletypesize{\scriptsize}
\tablecolumns{5}
\tablewidth{0pt}
\tablecaption{VLA Observations of 3C\,338.\label{VLA}}
\tablehead{\colhead{Date}& \colhead{Frequency} &\colhead{Configuration}&\colhead{Core flux density}&\colhead{rms} &\colhead{Total Flux}&\colhead{VLBI total\tablenotemark{*}}&\colhead{Peak Flux}\\
\colhead{}& \colhead{(GHz)} & \colhead{} &\colhead{(mJy)}&\colhead{(mJy/beam)} &\colhead{(mJy)}&\colhead{(\%)}& \colhead{mJy}}
\startdata

2002 Apr 12 &0.074&A &  ...     &    70 & 140900 &  ... &  32700 \\ 
2002 Apr 12 &0.330&A &  ...     &    1.1 &  21600 & ... &   900\\ 
1994 Nov 17 & 8.4 &C & 85    & 0.018 & 143  & 103  &  85\\
1995 Sep 11 & 8.4 &A & 88    & 0.027 & 103 & 98   &  88\\
1998 Dec 02 & 8.4 &C & 80    & 0.033 & 142  & 99   &  80\\
2000 Feb 26 & 8.4 &B & 82  & 0.026 & 132  & 110  &  82\\
2001 Jun 01 & 8.4 &BnC & 91 & 0.016 & 158 & 95 &  91\\

\enddata
\tablenotetext{*}{The VLBI total was obtained dividing the total cleaned flux in the VLBI observations at 8.4 GHz (see Table \ref{Observations}) by the VLA peak flux at 8.4 GHz}

\end{deluxetable*}

The 8.4 GHz and 15.4 GHz VLBI observations were carried out over the
course of six epochs, performed on 1994 Nov. 17, 1995 Sept. 11, 1997
Sept. 26, 1998 Dec 02, 2000 Feb. 26, and 2001 Jun 01 (see Table \ref{Observations}).
Observations in 1994, and 1995, were obtained using all ten elements
of the VLBA 
and a single VLA antenna.  Observations in 1997 onwards
were taken using a global array of between 12 and 15 antennas.  Both
right and left circular polarizations were recorded using 2 bit
sampling across a bandwidth of 8-16 MHz.  The VLBA correlator produced
16 frequency channels across each 8 MHz IF during every 2 s
integration.

\begin{deluxetable*}{lcccccccc}
\tabletypesize{\scriptsize}
\tablecolumns{8}
\tablewidth{0pt}
\tablecaption{VLBI Observational Parameters for 3C\,338 \tablenotemark{*}.\label{Observations}}
\tablehead{\colhead{Date}&\colhead{Frequency}&\colhead{Bandwidth}
&\colhead{Antenna}&\colhead{Scan}&\colhead{Time}&\colhead{Peak}
&\colhead{rms} & \colhead{Total Cleaned Flux}\\
\colhead{} &\colhead{(GHz)}&\colhead{(MHz)} &\colhead{} 
&\colhead{min}&\colhead{hours}&\colhead{mJy}&\colhead{mJy/beam} & \colhead{(mJy)}}
\startdata

1994 Nov 17 & 8.4 & 15 & VLBA+Y1 & 60 & 7.1 & 29 & 0.101 & 88 \\
1995 Sep 11 & 8.4 & 14 & VLBA+Y1 & 60 & 7.1 & 22 & 0.088 & 87 \\
1997 Sep 26 & 8.4 & 28 & VLBA+Y1+EB  & 60 & 9.9 & 18 & 0.033 & 72 \\
1998 Dec 02 & 8.4 & 32 & Global \tablenotemark{1}& 24 & 4.8 & 21 & 0.029 & 79 \\
           & 15.4 & 32 & VLBA+Y27+EB & 24 & 4.7 & 18 & 0.114 & 50 \\
2000 Feb 26 & 8.4 & 32 & Global \tablenotemark{2} & 20 & 5.0 & 27 & 0.037 & 91 \\
           & 15.4 & 32 & VLBA+Y27+EB & 20 & 4.0 & 25 & 0.133 & 71\\
2001 Jun 01 & 8.4 & 32 & Global \tablenotemark{3}& 30 & 5.0 & 37 & 0.045 & 86 \\
           & 15.4 & 32 & VLBA+Y27+EB & 30 & 5.5 & 22 & 0.114 & 51 \\

\enddata
\tablenotetext{*}{Global = VLBA + EB, MC, WB, GO, RO, Y27. Telescope Codes: VLBA= Very Long Baseline Array, VLA = Very Large Array, Y1 = single VLA antenna, Y27 = 27 VLA antennas, EB = Effelsberg, MC = Medicina, WB = Westerbork, GO = Goldstone, RO = Robledo.}
\tablenotetext{1}{GO, RO, and WB were out, as well as the VLBA antenna in Saint Croix.}
\tablenotetext{2}{GO, and RO were out, as well as the VLBA antenna in Hancock.}
\tablenotetext{3}{WB was out.}

\end{deluxetable*}

Parallactic angle effects resulting from the altitude-azimuth antenna
mounts were removed using the AIPS task CLCOR.  Amplitude calibration
for each antenna was derived from measurements of antenna gain and
system temperatures during each run.  Delays between the stations'
clocks were determined from a short observation of the bright
calibrator 3C\,279 using the AIPS task FRING \citep{Schwab83}.
Calibration was applied by splitting the multi-source data set
immediately prior to preliminary editing, imaging, deconvolution, and
self-calibration in {Difmap} (Pearson et al. 1994).  Multiple
iterations of phase self-calibration and imaging were applied to each
source (including calibrators and 3C\,338) before any attempt at
amplitude self-calibration was made.  The preliminary models developed
in {Difmap} were subsequently applied in AIPS to refine the gain
corrections, to determine the leakage terms between the RCP and LCP
feeds (using the unpolarized calibrator OQ\,208) and to correct for
residual phase differences between polarizations.  Final imaging and
self-calibration were performed in {Difmap}.

No polarized flux has been detected from 3C\,338 at either
8.4 or 15.4 GHz in epochs 2000 or 2001 in which careful polarization
calibration was performed.  Typical 2$\sigma$ limits on the
linearly polarized flux density are $<$70 $\mu$Jy at 8.4 GHz and
$<$200 $\mu$Jy at 15.4 GHz.
This corresponds to a limit of $<0.4\%$ for the jet component E2.
The low polarization is consistent with the high rotation measure
expected for an orientation of the inner radio jets close to the plane of the sky (Zavala \& Taylor 2002).

\begin{figure*}
\centering
\scalebox{0.5}{\includegraphics[angle=0]{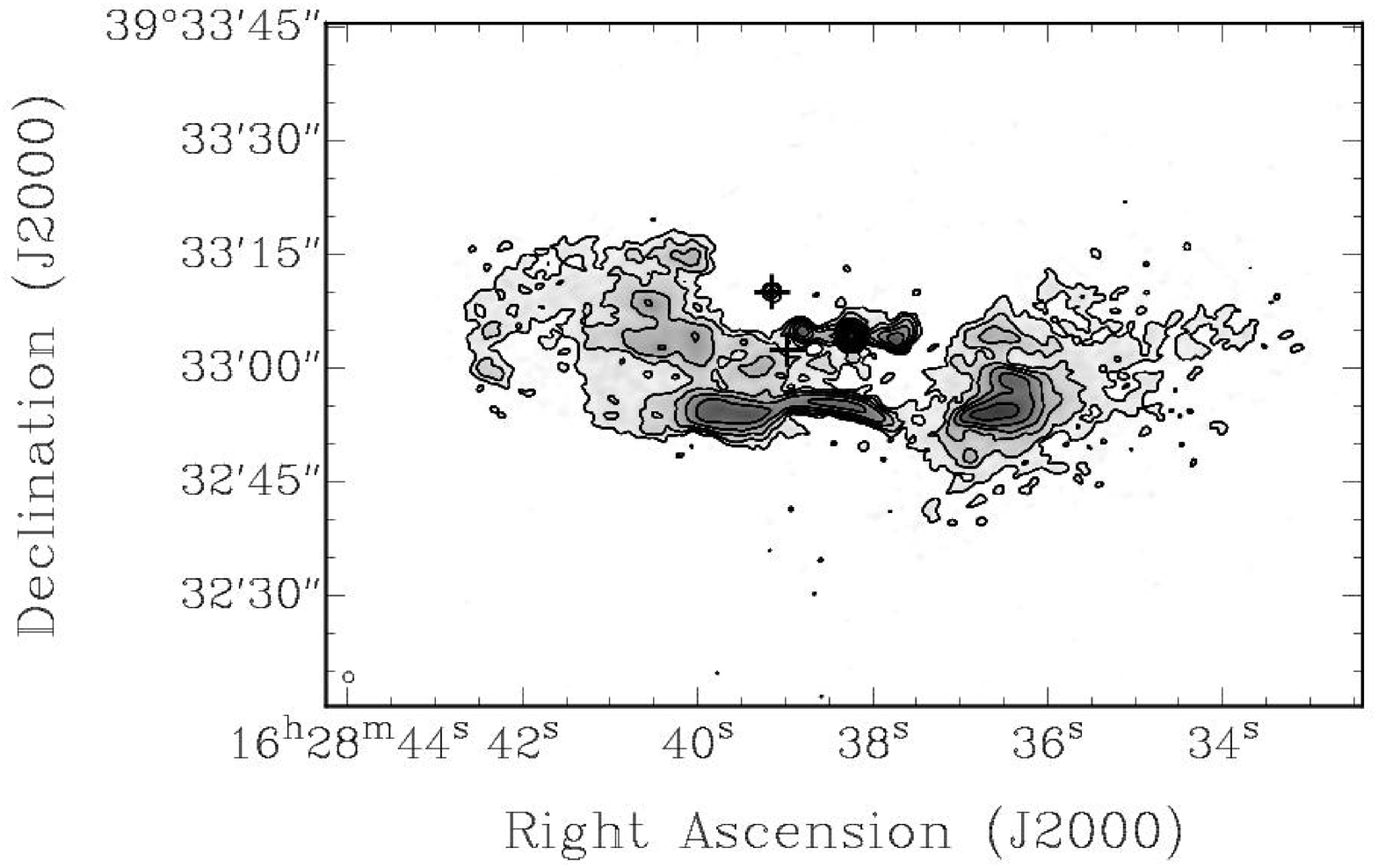}}
\caption{Contours of the 8.4 GHz map of 3C\,338 from the VLA A, B, BnC, and C configurations at 1.4$^{\prime\prime}$ resolution. The synthesised beam is shown in the bottom left corner.
Contours levels are 32 (4$\sigma$), 64, 128, 256, ... $\mu$Jy beam$^{-1}$.
The crosses indicate the location of component B (north) and 
C (south); see Section \ref{vlaimages}.
}
\label{vla_full}
\end{figure*}


\section{Results}
\label{results}

\subsection{VLA Images}
\label{vlaimages}

The low frequency maps are shown in Figures \ref{74} and \ref{330}. The emission at those frequencies is more extended than 
the higher frequency maps presented in \cite{Giovannini98} and \cite{Johnstone02}, especially in the western lobe, even though the comparison at 74 MHz  
is difficult due to the relatively poor angular resolution.
An example of a new feature in the 330 MHz, compared to the higher  
frequency maps, is the plume extending northwards
from the northern tip of the western lobe; this feature evident, but somewhat smeared  
out at 74 MHz. The source total flux is
reported in Table \ref{VLA}. We estimated also the flux density for the west lobe 
(67.7 Jy at 74 MHz and 10.3 Jy at 330 MHz) and for the east lobe (76.2 Jy
at 74 MHz and 11.3 Jy at 330 MHz) by integrating over regions defined by the lobe
shapes; since the core is relatively weak at those frequencies, it is not crucial
to select carefully the regions over which the integration has to be done.

Figure \ref{vla_full} shows the VLA image at 8.4 GHz and resolution of 1.4$^{\prime\prime}$,  
from A, B, BnC, and C configurations, obtained by combining the visibilities from the five epochs;
the map was deconvolved using a multi-resolution CLEAN (Wakker \& Schwarz 1988), to highlight both 
the small-scale and the large-scale structures. This algorithm
is a variant of the standard CLEAN (H\"ogbom 1974), which
uses Gaussian components of different sizes; in our case we used components
with sizes up to 16\arcsec.
The total flux in this map is 157.8 mJy, with a peak of 86.1 mJy beam$^{-1}$.
The west and east lobe flux density is respectively 44 mJy, and 43 mJy; 
the difference with respect to the total flux density is due to the core and 
central jet region. 
The structure observed in 3C\,338 is similar to the one obtained by \cite{Ge94} at 5 GHz, with two diametrically opposed jets, as well as a peculiar ridge connecting the eastern and western
lobes of emission. The orientation and position of this image is
consistent with our VLA images at 74 and 330 MHz. 
The optical center of the cluster consists of four components (A, B, C, and D, Minkowski 1961; 
Burbidge 1962), the brightest of which corresponds to the core of the
radio emission (Burns et al. 1983; Fanti et al. 1986).
Recent HST observations (Martel et al. 1999, Capetti et al. 2000) have found the second and
third brightest components (B and C) to be elliptical galaxies at similar projected distances
from NGC 6166. Their positions are indicated in Figure \ref{vla_full}.
\cite{Ge94} identified a weak feature found in their 5 
GHz maps, to the northeast of the radio core, as optical component B. This component
is also present in our 8.4 GHz map, with a peak flux of 0.15 mJy beam$^{-1}$.

\subsection{VLBI Images}

\begin{figure}
\centering
\includegraphics[width=8.5cm]{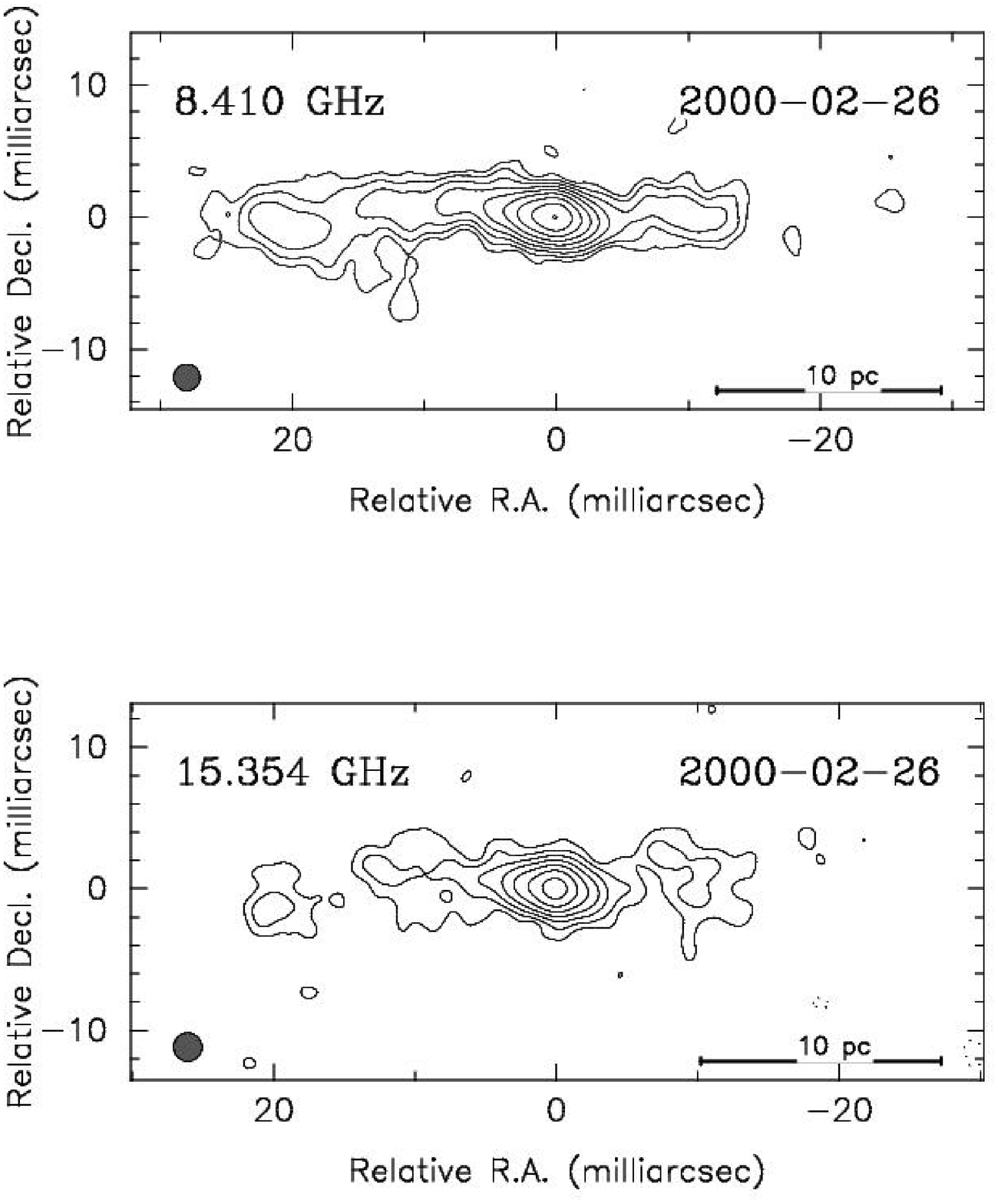}
\caption{Contemporaneous VLBA observations of 3C\,338 at 
8 and 15 GHz from the 2000 epoch.  Both images have been 
tapered and restored with a circular 2 mas
synthesized beam (drawn in the lower left-hand corner
each plot).  Contours are drawn logarithmically at factor 2 intervals with 
the first contour at 0.15, and 0.4 mJy/beam at 8 and 15 GHz 
respectively.}
\label{best}
\end{figure}

Figure \ref{best} shows ``naturally'' weighted 8.4 and 15.4 GHz images from the VLBI 2000 observations. ``Natural'' weighting does not correct
for the local density of samples in the $(u,v)$ plane and thus gives a lower resolution.  
Both images were tapered and restored with a circular 2 mas synthesized beam. 
The orientation of our VLBI images is consistent with the orientation 
observed in our VLA images at 0.074, 0.330, and 8.4 GHz. 
VLBI observations recover 95\% or more of the VLA core flux density at 
8.4 GHz (Table \ref{VLA}).
 In 2000 the VLBI total is 10\% higher than the VLA, indicating a calibration 
error of $\sim 10\%$ in one or both data sets.

\begin{figure*}
\centering
\scalebox{0.75}{\includegraphics{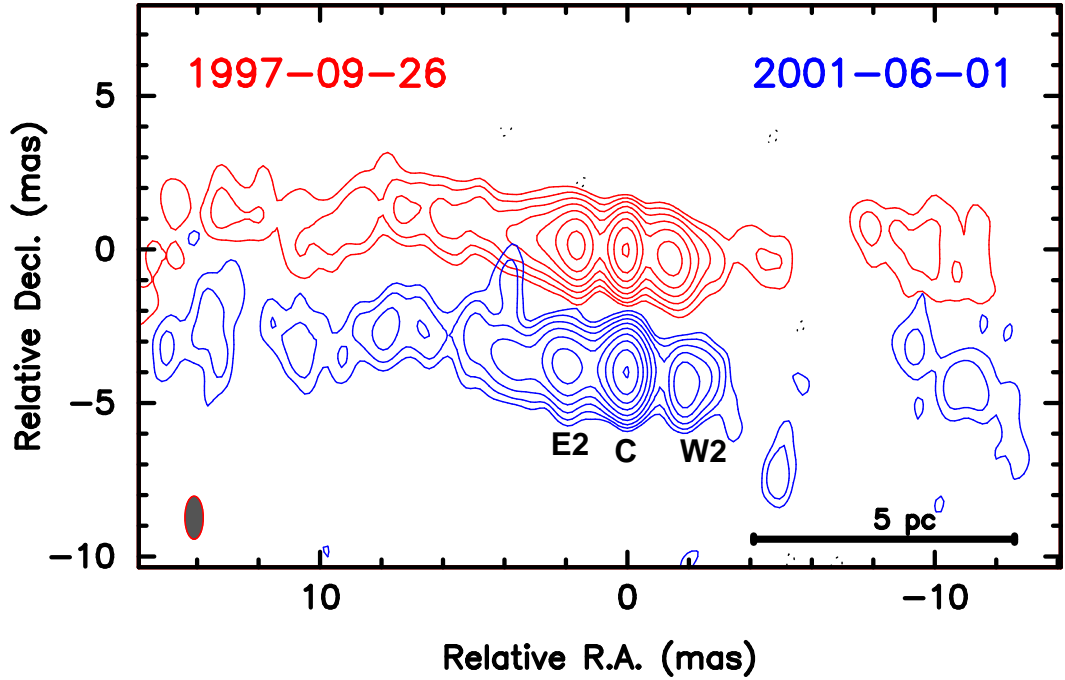}}
\caption{An overlay at 8.4 GHz of the 1997 (in red) and 2001 (in blue) VLBA observations at
full resolution of 1.4 $\times$ 0.6 mas in position angle 0\deg.  Contours are drawn starting at 0.13 mJy/beam for both epochs and increase by factors of two thereafter.}
\label{red-blue}
\end{figure*}

In Figure \ref{red-blue}, which shows the 8.4 GHz 1997 and 2001 epochs at 
full resolution of 1.4 $\times$ 0.6 mas (PA 0$^{\circ}$), we see a bright 
central component, which we designated C, as well as two jets to the 
east and west respectively. We designated the brightest component in the 
eastern jet as E2, and the brightest component in the western jet as W2. 

Motion and variability studies were performed by fitting six elliptical  or circular Gaussian components in Difmap to the 2000 self-calibrated visibility data at 8.4 GHz. Then, this model was used to fit the data corresponding to 1994, 1995, 1997, 1998, and 2001 epochs at the same frequency. We let only position and flux density vary, in order to fit the independently self-calibrated visbility data, while all the other parameters were held fixed at the 2000 values. Results from our fits are listed in Table \ref{Big_Gaussian}.
Since phase-referencing was not performed, all positions are 
relative positions only.  We selected the strong compact, VLBI
core (component C) as the reference.  Statistical errors in the
positions range from 0.02 to 0.06 mas and are derived from 
the component size divided by the signal-to-noise ratio.

\begin{deluxetable*}{lcccccccc}
\tabletypesize{\scriptsize}
\tablecolumns{9}
\tablewidth{0pt}
\tablecaption{Gaussian Model Components\tablenotemark{*}.\label{Big_Gaussian}}
\tablehead{\colhead{Component}&\colhead{Epoch}&\colhead{$S$}&\colhead{$r$}
&\colhead{$\theta$}&\colhead{$a$}&\colhead{$b/a$}&\colhead{$\Phi$}
&\colhead{$\chi^{2}$} \\
\colhead{} & \colhead{} & \colhead{(Jy)} & \colhead{(mas)} 
& \colhead{($^o$)}&\colhead{(mas)}&\colhead{}&\colhead{($^o$)}}
\startdata
C...  & 1994.877 & 0.022 $\pm$ 0.001 &  0.00     &     0.0 & 0.29 & 1.00 & 2.9 & 1.03  \\
      & 1995.693 & 0.016 $\pm$ 0.001 &  0.00     &     0.0 & 0.29 & 1.00 & 2.9 & 0.14  \\
      & 1997.741 & 0.018 $\pm$ 0.001 &  0.00     &     0.0 & 0.29 & 1.00 & 2.9 & 1.86  \\
      & 1998.916 & 0.022 $\pm$ 0.001 &  0.00     &     0.0 & 0.29 & 1.00 & 2.9 & 1.49  \\
      & 2000.151 & 0.029 $\pm$ 0.002 &  0.00     &     0.0 & 0.29 & 1.00 & 2.9 & 1.13  \\
      & 2001.415 & 0.033 $\pm$ 0.002 &  0.00     &     0.0 & 0.29 & 1.00 & 2.9 & 1.17  \\
E1... & 1994.877 & 0.012 $\pm$ 0.001 &  12.47   & 88.1    & 19.46 & 0.14 & 277.4 & 1.03  \\
      & 1995.693 & 0.015 $\pm$ 0.001 &  11.36   & 86.9    & 19.46 & 0.14 & 277.4 & 0.14  \\
      & 1997.741 & 0.012 $\pm$ 0.001 &  12.96   & 87.7    & 19.46 & 0.14 & 277.4 & 1.86  \\
      & 1998.916 & 0.013 $\pm$ 0.001 &  13.64   & 87.5    & 19.46 & 0.14 & 277.4 & 1.49  \\
      & 2000.151 & 0.016 $\pm$ 0.001 &  13.51   & 88.3    & 19.46 & 0.14 & 277.4 & 1.13  \\
      & 2001.415 & 0.010 $\pm$ 0.001 &  12.59   & 86.7    & 19.46 & 0.14 & 277.4 & 1.17  \\
E2... & 1994.877 & 0.023 $\pm$ 0.001 &  1.48    & 83.7    & 0.80  & 0.33 & 298.1 & 1.03   \\
      & 1995.693 & 0.016 $\pm$ 0.001 &  1.54    & 85.5    & 0.80  & 0.33 & 298.1 & 0.14   \\
      & 1997.741 & 0.012 $\pm$ 0.001 &  1.60    & 85.8    & 0.80  & 0.33 & 298.1 & 1.86   \\
      & 1998.916 & 0.010 $\pm$ 0.001 &  1.71    & 81.8    & 0.80  & 0.33 & 298.1 & 1.49   \\
      & 2000.151 & 0.008 $\pm$ 0.001 &  1.76    & 84.9    & 0.80  & 0.33 & 298.1 & 1.13   \\
      & 2001.415 & 0.0040 $\pm$ 0.0002 &  1.92    & 89.7    & 0.80  & 0.33 & 298.1 & 1.17   \\
E3... & 1994.877 & 0.018 $\pm$ 0.001 &  1.03    & 81.7    & 5.33  & 0.06 & 80.3  & 1.03   \\
      & 1995.693 & 0.027 $\pm$ 0.002 &  0.63    & 72.0    & 5.33  & 0.06 & 80.3  & 0.14   \\
      & 1997.741 & 0.019 $\pm$ 0.001 &  1.60    & 82.6    & 5.33  & 0.06 & 80.3  & 1.86   \\
      & 1998.916 & 0.022 $\pm$ 0.001 &  1.35    & 88.3    & 5.33  & 0.06 & 80.3  & 1.49   \\
      & 2000.151 & 0.026 $\pm$ 0.001 &  1.23    & 80.7    & 5.33  & 0.06 & 80.3  & 1.13   \\
      & 2001.415 & 0.023 $\pm$ 0.001 &  1.63    & 82.0    & 5.33  & 0.06 & 80.3  & 1.17   \\
E4    & 2001.415 & 0.009 $\pm$ 0.001 &  0.38    & 80.1    & 0.48  & 1.00 & 353.3 & 1.17 \\
W1... & 1994.877 & 0.008 $\pm$ 0.001 &  8.85    & 271.2   & 18.39 & 0.13 & 275.6 & 1.03   \\
      & 1995.693 & 0.008 $\pm$ 0.001 &  6.05    & 266.2   & 18.39 & 0.13 & 275.6 & 0.14   \\
      & 1997.741 & 0.0050 $\pm$ 0.0003 &  8.31    & 269.8   & 18.39 & 0.13 & 275.6 & 1.86   \\
      & 1998.916 & 0.0070 $\pm$ 0.0004 &  7.81    & 268.4   & 18.39 & 0.13 & 275.6 & 1.49   \\
      & 2000.151 & 0.009 $\pm$ 0.001 &  6.88    & 268.7   & 18.39 & 0.13 & 275.6 & 1.13   \\
      & 2001.415 & 0.0030 $\pm$ 0.0002 &  10.08   & 266.7   & 18.39 & 0.13 & 275.6 & 1.17   \\
W2... & 1994.877 & 0.007 $\pm$ 0.001 &  1.25    & 255.6   & 0.85  & 0.47 & 53.1  & 1.03   \\
      & 1995.693 & 0.0050 $\pm$ 0.0003 &  1.30    & 255.1   & 0.85  & 0.47 & 53.1  & 0.14   \\
      & 1997.741 & 0.0080 $\pm$ 0.0004 &  1.48    & 260.5   & 0.85  & 0.47 & 53.1  & 1.86   \\
      & 1998.916 & 0.0050 $\pm$ 0.0003 &  1.71    & 261.0   & 0.85  & 0.47 & 53.1  & 1.49   \\
      & 2000.151 & 0.0040 $\pm$ 0.0002 &  1.82    & 254.7   & 0.85  & 0.47 & 53.1  & 1.13   \\
      & 2001.415 & 0.0040 $\pm$ 0.0003 &  1.99    & 259.1   & 0.85  & 0.47 & 53.1  & 1.17   \\

\enddata
\tablenotetext{*}{NOTE - Parameters of each Gaussian component of the model
brightness distribution are as follows:  Component, Gaussian component; 
Epoch, year of observation (see Table \ref{Observations});  
$S$, flux density; $r$, $\theta$, polar coordinates of the
center of the component relative to the center of component C 
; $a$, semimajor axis; $b/a$, axial ratio; 
$\Phi$, component orientation; $\chi^{2}$, goodness-of-fit for six component model in each epoch. 
All angles are measured
from north through east.  Errors in flux are based on our absolute amplitude calibration
as well as the rms noise.
See text for a discussion of the position errors.}

\end{deluxetable*}

We obtained an estimate for the sizes of C, E2, and W2 based on 
our 8.4 GHz model fits (see Table \ref{Big_Gaussian}), 
finding that they are all compact with sizes smaller than 0.58 mas or 0.36 pc.
To fit the data two more components were necessary: component E1 (extended
and diffuse at large distance from component C) to recover the extended jet flux
density and component E3 to fit the jet in between components C and E2.

A new component (E4) was found in the 2001 VLBI  observations at 8.4 GHz.
This component is at $\sim 0.4$ mas east from component C and is almost twice as large as it, implying that new jet components are still being ejected from the core. This component is however always visible in 15 GHz images and probably 
not seen in 1998 and 2000 data at 8.4 GHz because of its more limited angular resolution.

\begin{figure*}
\centering
\scalebox{0.6}{\includegraphics{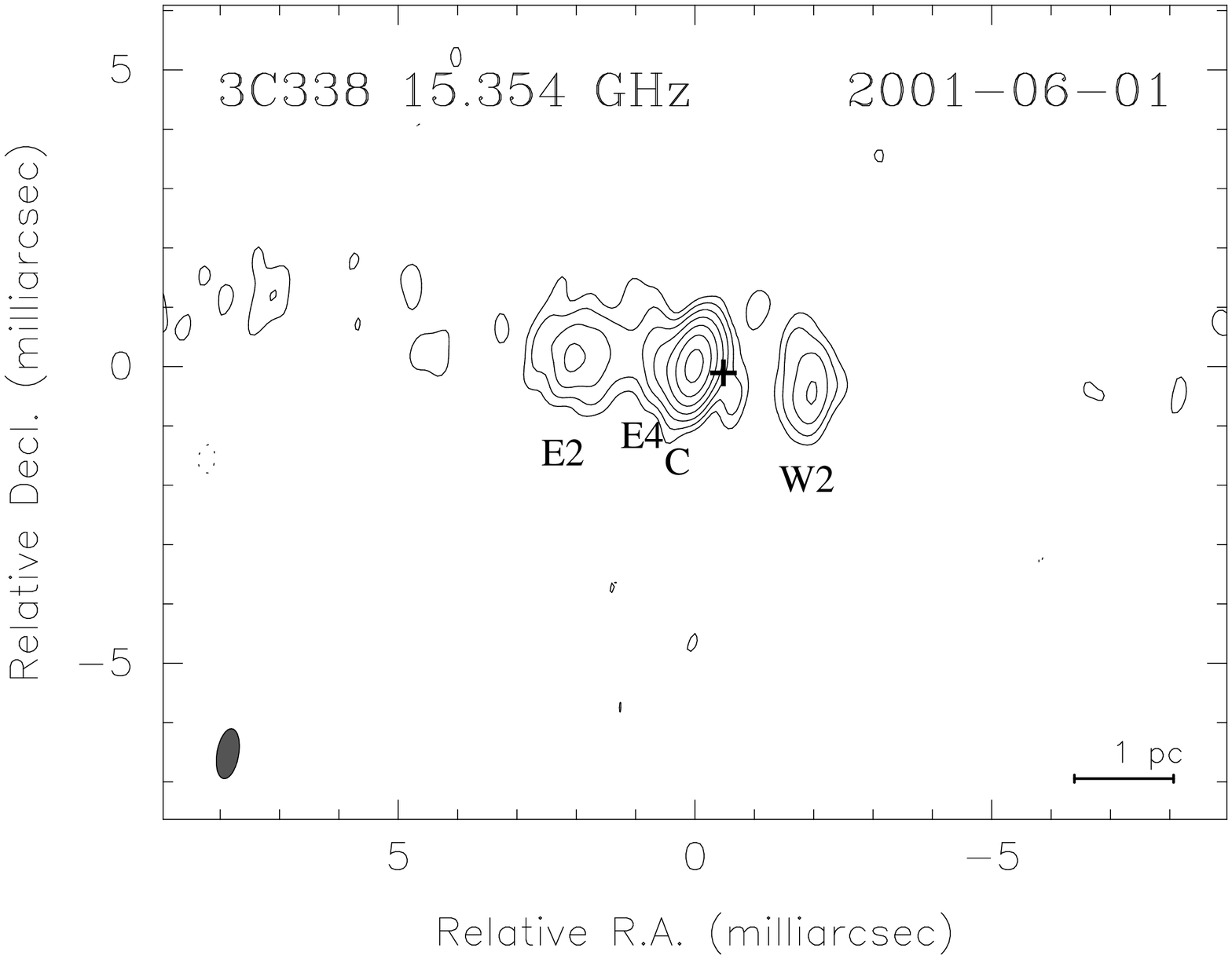}}
\caption{
15 GHz VLBI map at full resolution, using natural weighting. Contours start at 2.5$\times 10^{-4}$ mJy beam$^{-1}$.
The beam is 0.85 $\times$ 0.37 mas in position angle -9.7\deg. The components fitted from the 8 GHz data are shown;
the cross indicates the position of the origin of the jets in our model.
}
\label{vlbi_15}
\end{figure*}

For the model-fitting
we used the 8.4 GHz data instead of those at 15 GHz 
(shown in Figure \ref{vlbi_15}) 
because in the latter the noise is 3 times higher while the 
components are half the flux density. Hence, despite their higher resolution, 
the 15 GHz data provide weaker constraints; however the model-fitting 
performed on the 15 GHz shows the same trend as the 8.4 GHz data.

\subsection{Motions and Variability}
\label{motion_variability}

We do not discuss components E1 and E3 since these two last components 
are extended and diffuse and do not correspond to well defined structure.
Component E4 is present in only one epoch at 8.4 GHz, but it is visible in all
epochs at 15 GHz with no appreciable proper motion.

To calculate the relative velocity of the components, we chose component C as the reference. We were able to compare the relative motions of both E2 and W2 by fitting a line to each component's projected separation, derived from the Gaussian modelfitting, as a function of time. The results obtained are shown in Table \ref{Motions_Table} and plotted in Figure \ref{velplot}. Component E2 is moving away from component C to the east, while component W2 is moving away from it to the west. 

Significant motion with respect to component C was obtained for both E2 and W2, yielding a value of $0.055 \pm 0.005$ mas/y or ($0.112 \pm 0.010$)c and $0.115 \pm 0.008$ mas/y or (0.234 $\pm$ 0.016)c respectively.

\begin{deluxetable*}{lccc}
\tabletypesize{\scriptsize}
\tablecolumns{8}
\tablewidth{0pt}
\tablecaption{Component Motion Fitting Results.\label{Motions_Table}}
\tablehead{\colhead{Component}&\colhead{Velocity}&\colhead{Velocity}&\colhead{Angle of Motion}\tablenotemark{*} \\
\colhead{}&\colhead{(mas/y)}&\colhead{(c)}&\colhead{($^o$)}}
\startdata
C  & Reference Component &... &...  \\
E2 &  0.055 $\pm$ 0.005 & 0.112 $\pm$ 0.010 & 90.84  \\
W2 &  0.115 $\pm$ 0.008 & 0.234 $\pm$ 0.016 & 264.44  \\
\enddata
\tablenotetext{*}{Angles measured from north through east.}
\end{deluxetable*}

To study component variability in 3C\,338, we compared the flux density for components C, E2, and W2 over each of our six epochs at 8.4 GHz. Errors for each region were calculated based on the rms noise and our estimated absolute flux calibration errors ($\sim 5 \%$). The resulting fractional variation light curves are shown in Figure \ref{lightcurves}. 
We also added the core's VLA flux density light curve found from the five epochs studied at 8.4 GHz. Errors for this region were based on our estimated absolute flux calibration errors ($\sim 3 \%$).
These light curves were created by dividing each region's flux density at each epoch by the flux density measured in the first epoch. The light curves for components E2 and W2, and the core's VLA flux density are displaced on the $y$-axis by 1, 2, and 3 units, respectively.

From Figure \ref{lightcurves} it can be noted that from 1994 to 1995 the displayed light curves for components C, E2, and W2 decreased in flux density, independently of the subsequent evolution. This leads us to believe that there was an error of about 30\% in the calibration of the 1994 data. Component C substantially increases in flux, going from 16 mJy in 1995 to 33 mJy in 2001. Component E2, on the other hand, decreases in flux density, going from 16 mJy in 1995 to 4 mJy in 2001. Finally, component W2 is less monotonic, since it increases in flux density from 1995 to 1997 (going from 5 mJy to 8 mJy), and then it decreases from 1997 to 2001 (going from 8 mJy to 4 mJy). 
Finally, we find that there is no significant variation in the
VLA core flux density light curve over the 7 yr baseline.
This lack of variability was unexpected from previous observations (Giovannini et al. 1998), since
the arcsecond core flux density was a factor $\sim$ $1.5 - 2$ higher in 1990-1991. However, more recent
8 GHz VLA data from 2005 (Partridge \& Lin, priv. comm.) indicate a core flux density of 84 mJy, consistent with
the very small variations found in $1994 - 2001$. It appears that the core has been in a quiescent phase since
1994.

\subsection{Spectral Index Distribution}

The 1998 VLBI images at 8.4 and 15.4 GHz were matched in
resolution in order to obtain a spectral index distribution across the
source (Figure \ref{spix}). This epoch was chosen based on the quality of the data, number
of antennas out during observations, and noise in the final image. 
Table \ref{Peak_Fluxes} lists the flux densities measured from
the matching resolution images, as well as the spectral index obtained 
for components C, E2, and W2.
A steep
spectrum was found for both the eastern and western jets, obtaining a spectral index equal
to $- 0.95$ for E2, and $- 0.71$ for W2.
The region near E2 and W2 components appears flatter than  
connecting regions. The very inverted (red in Figure \ref{spix}) regions could suggest the presence
of a jet structure or are probably due to the different uv-coverage between the
two data sets.
Component C has a flatter spectral index of -0.11, indicating that it is associated 
with the center of activity (see Table \ref{spectra})

At arcsecond resolution we did not derive a spectral index image because
of the low resolution of VLA 74 MHz data. The integral spectral index
for the two lobes is $\alpha^{74}_{330}$ = -1.26 and $\alpha^{0.3}_{8.4}$
= -1.68 for the west lobe and $\alpha^{74}_{330}$ = -1.28 and 
$\alpha^{0.3}_{8.4}$ = -1.72 in agreement with previous results 
(Burns et al. 1983).


\section{Discussion}

\subsection{Kinematics in 3C\,338 and Evolution of Jet Components}

The symmetric source structure on the parsec scale suggests an orientation near 
to the plane of the sky. However the jet
length and total flux density (see Figs. 5 and 6) and the asymmetric E4 
component suggests that the east side is the approaching one. Therefore
we will consider in the next discussion the east jet as the main jet and the
west one as the counter-jet.

\subsubsection{Jet orientation and bulk velocity of 3C\,338 on the parsec scale}

In the simple beaming model for simultaneously ejected jet components moving in opposite
directions, either the arm length ratio $ D $ or the flux density ratio
$ R $ can be used as a first constraint on the intrinsic speed $\beta=v/{\rm c}$ and the angle of the
jets to the line of sight $\theta$
\citep{Taylor&Vermeulen97}. The arm length ratio, $D$, is given by

\begin{equation}
\label{arm_ratio}
D=\frac{d_{E}}{d_{W}}=\left( \frac{1+\beta \cos \theta}{1-\beta \cos \theta}\right),
\end{equation}

\noindent
where the apparent projected distances from C (assumed for the moment to be
the origin of the jet)
are $ d_{E} $ for the eastern jet (approaching side) and $ d_{ W} $ for 
the western jet (receding side).

We applied this relation to the whole parsec scale structure visible
in the 8.4 GHz images (e.g. Fig. 4a, top) and we measured an arm ratio
$\sim$ 1.87 corresponding to
$\beta {\rm cos}\theta$ $\sim$ 0.30.
Alternatively we could consider the distance from the core of components
E2 and W2 but because of different proper motion velocity we have
an arm length ratio from 1.18 (1994 epoch)to 0.96 (2001 epoch).

Similarly, the flux density ratio, $R$, between the eastern and western jet is

\begin{equation}
\label{flux_ratio}
R=\frac{S_{E}}{S_{ W}}=\left( \frac{1+\beta \cos\theta}{1-\beta \cos\theta}\right)^{k-\alpha}
\end{equation}

\noindent
where $ \alpha $ is the spectral index, $ k = 2 $ for a continuous jet, and 
$ k = 3 $ for discrete jet components (e.g. Lind \& Blandford 1985).

The flux density ratio was obtained dividing the sum of the fluxes of 
component E2 over the six epochs by the sum of the fluxes of component W2 
over the six epochs (see Table \ref{Big_Gaussian}), yielding a value of 
$ R=S_{E}/ S_{W} = 2.21 \pm 0.33$. From this result we can obtain a value 
for $\beta {\rm cos} \theta$. Using $k = 2$ and the mean spectral index 
between E2 and W2, $\alpha = -0.83 \pm 0.06$, we obtained 
$\beta {\rm cos} \theta = 0.14 \pm 0.04$

If we consider component E4 and assume that the symmetric W4 component is
missing because of relativistic effects on a symmetric emission, we derive
a brightness ratio R $>$ 12 corresponding to $\beta {\rm cos} \theta$ 
$>$ 0.46.

\subsubsection{Jet orientation and velocity of 3C\,338 on the parsec scale from
proper motion measures}

We can also constrain the jet orientation and velocity (pattern 
velocity) by comparing different epoch data and measuring the proper motion
of jet substructures.

\begin{figure*}
\centering
\scalebox{0.33}{\includegraphics{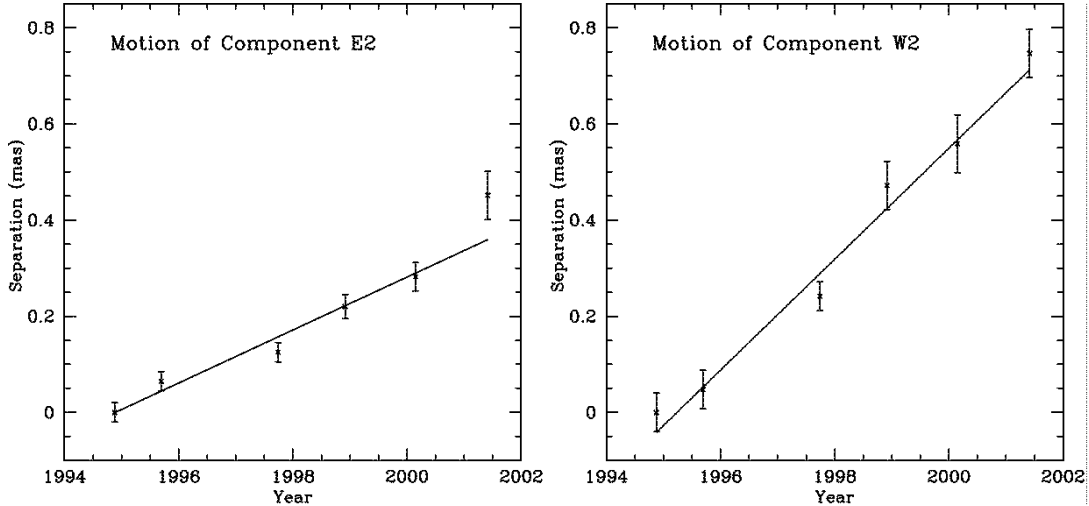}}
\caption{Velocities of the components E2 and W2 derived from the 
straight line which best fits the
measured positions from the 8.4 GHz data only (see Fig.~3).
The zero point is taken to be the position of the component at
the first epoch. See text for a discussion of the position errors.}
\label{velplot}
\end{figure*}

\begin{figure}
\centering
\scalebox{0.4}{\includegraphics{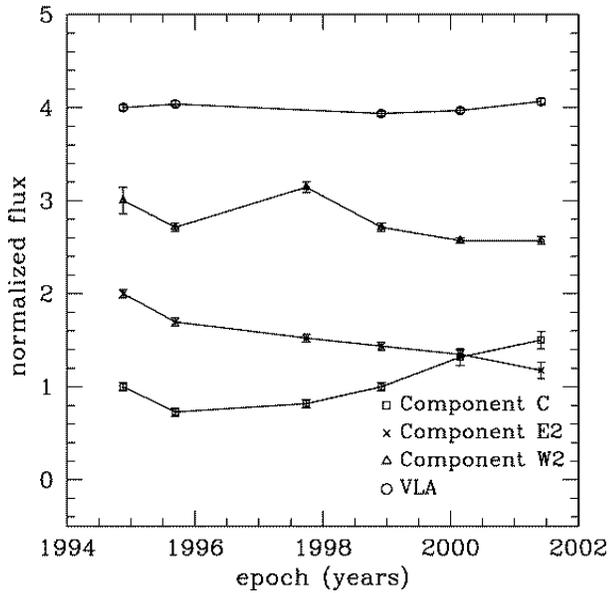}}
\caption{Light curves of the three brightest components of 3C\,338: C, E2, and W2, as well as the total VLA flux density. The values used to produce
this graph were taken from Tables \ref{Big_Gaussian} (for C, E2, and W2) and \ref{VLA} (for the core's VLA peak flux) and are discussed in \S \ref{motion_variability}. The displayed light curves were created by dividing each region's flux at each epoch by the first epoch's flux of each region respectively. The light curves of component E2, component W2, and the core's VLA peak flux are displaced on the $y$-axis by 1, 2, and 3 units respectively. Errors are estimated from the rms noise and the absoulute flux calibration errors for each epoch.}

\label{lightcurves}
\end{figure}

\begin{figure}
\centering
\scalebox{0.38}{\includegraphics{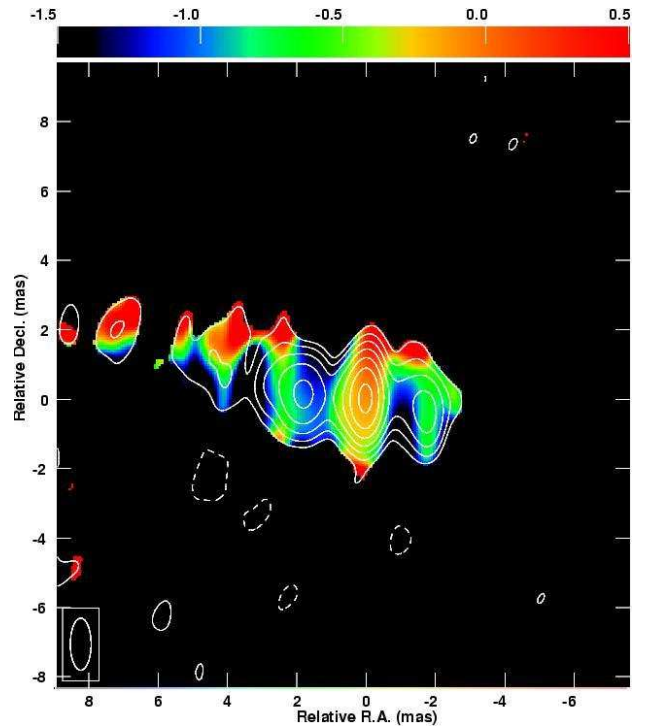}}
\caption{Spectral index distribution between 8.4 and 15.4 GHz from the
1998 observations. The contours are taken from the 15.4 GHz observations and are
set at $3 \sigma $, increasing by a factor of 2 thereafter.}
\label{spix}
\end{figure}

In Figure \ref{velplot} we plot the positions of the two strongest jet 
components, E2 and W2, derived from the modelfitting.  These inner jet 
components both appear fairly straight to within the errors in the 
measurements: there is no clear sign of acceleration or deceleration.
Given these motions and the symmetry of the source (Figures \ref{best} 
and \ref{red-blue}), it
is reasonable to assert that E2 and W2 were ejected at the same time.
Deep observations at 8.4 GHz reveal that
there is continuous emission from the core outwards.  The
jet is probably not made up of discrete blobs that can be well
described by elliptical Gaussian components that we identify and
modelfit.  Rather the jet appears to be a continuous flow, with
features of enhanced emission (possibly shocks)  that propagate down the jet.

Using the values obtained for the velocities of components E2 and W2 (see Table 
\ref{Motions_Table}), and using the distance from these components to 
component C 
(see Table \ref{Big_Gaussian}) 
we calculated the velocity ratio obtaining
$\mu _E / \mu _W = 0.478 \pm 0.077$ 
and $d_E / d_W = 0.965 \pm 0.001$ respectively. 

In this case the approaching side of the jet is 
moving slower than the receding side.
This result either i) implies that the measured proper motion is not a real
motion but is due to variability of jet brightness or ii)  requires that 
the real core is moving to the west. 

The first case has been discussed in many sources where apparent motion
with different velocities as well as fixed structures have been found
as in M87 (see e.g. Dodson et al. 2006 and references therein), and Markarian 501 
(Giroletti et al. 2004).

In the second case
by assigning a velocity to component C we can make the velocity and  
arm length ratio consistent with the assertion that E2 and W2 were ejected at 
the same time, that is, we can make these two ratios equal. Under this 
assumption we obtain that component C is moving to the east at 
$\mu_c=0.058{\rm c}$. With this value we then determined the velocity of both 
E2 and W2 with respect to component C and finally estimate their age to be 
23 yr, meaning that they were ejected roughly in 1978.
This implies a total displacement of component C of 0.63 mas, or about one 
beam FWHM. The true core
could still be unseen if its flux is smaller than about 4\% of the peak flux 
of C, or about 1 mJy/beam.
We expect that in a few more years the core (which should be stationary) 
should become apparent, if this model is correct. Component C is 
required to be intrinsically brigther on one side, i.e. its corresponding
component west of the true core is required to be suppressed.

Another independent constraint on the two parameters $\beta$ and $\theta$ can be obtained from the separation rate $\mu _{sep} = \vert \mu _{E} \vert + \vert \mu _{W} \vert$, which is not subject to the uncertainty in the reference point. From geometry and the conversion of angular to linear velocity we have

\begin{equation}
\label{vsep}
v_{sep} = \mu _{sep} D_{a} (1 + z) = \frac{2~ \beta~ c ~\rm sin~ \theta}{(1 - \beta ^2 {\rm cos} ^2 \theta)},
\end{equation}

\noindent
where $v_{sep}$ is the projected separation velocity, $D_a$ is the angular size distance to the source, and $z$ is the redshift.
Using the values obtained for the motion of components E2 and W2, we obtained a projected velocity separation equal to $v_{sep} = (0.346 \pm 0.026)c$.

\subsubsection{Jet kinematics and orientation}

Observational results discussed above do not present strong constraints on the
jet kinematics and orientation for this symmetric source.
We have two different possible explanations for the parsec scale jet structure in 
3C\,338:

1) Jets on the parsec scale are only mildly relativistic and the pattern
velocity from the proper motion is indicative of the real jet velocity.
In this case the 3C\,338 nuclear source is moving as discussed above, and this 
could be in agreement with the nuclear dynamics of this complex cD galaxy
where we have 4 different optical nuclei: a strong 
evidence that this galaxy
suffered many major mergers and that the central region is not yet relaxed.

\begin{figure*}
\centering
\vspace{1cm}
\scalebox{0.7}{\includegraphics{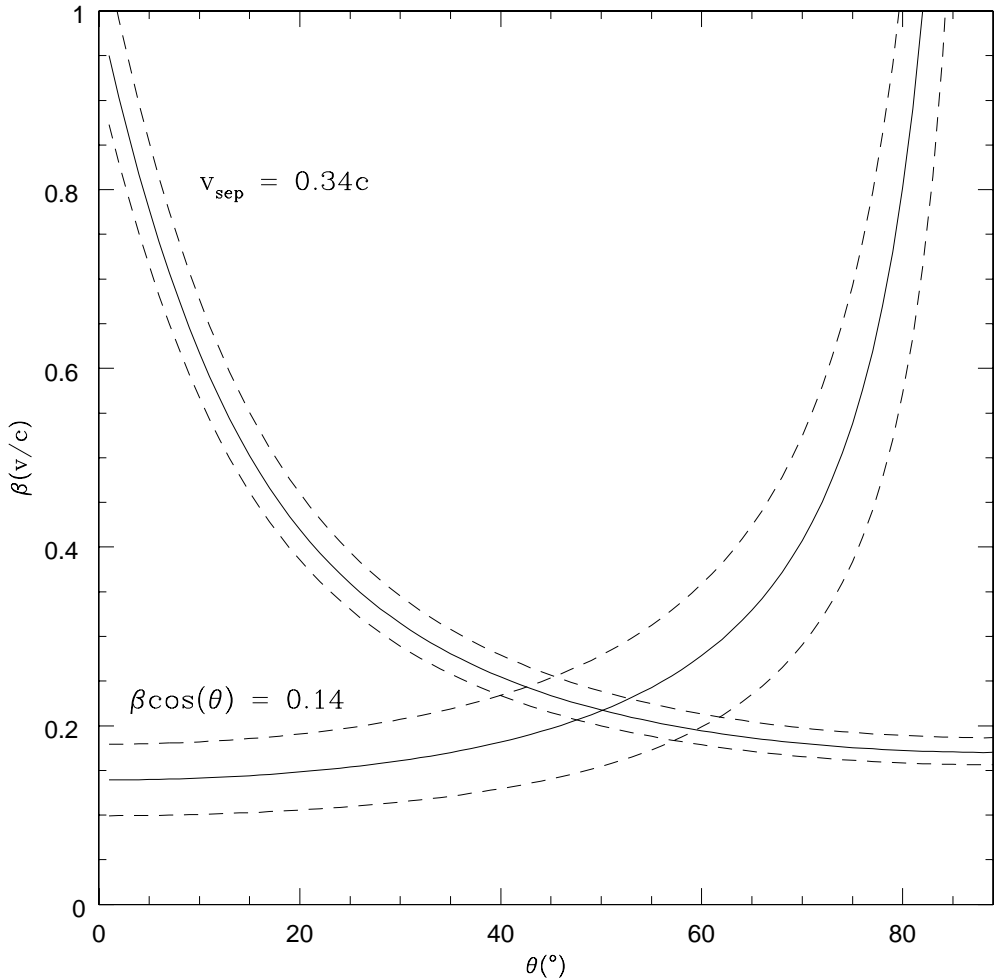}}
\caption{The jet velocity ($\beta$) plotted against the inclination
of the source ($\theta$) measured from the line of sight to the jet
axis.  The solid increasing line represent the constraint $\beta \cos\theta =
0.14 \pm 0.04$ from the flux ratio of components E2 and W2.
The solid decreasing line shows the constraint from the observed
separation velocity, $h^{-1} v_{sep}$, for E2 and W2 with $h$ =
0.71 $\pm$ 0.05 where $h = H_0/100$ km s$^{-1}$ Mpc$^{-1}$. 
The dashed lines represent the uncertainties.
}
\label{betathew}
\end{figure*}

Figure \ref{betathew} shows the jet velocity ($\beta$) plotted against 
the inclination of the source ($\theta$). The intersection obtained by 
plotting the results for $\beta cos \theta$ and $v_{sep}$ gives us 
constrains for both 
values $\beta$ and $\theta$. For $\beta cos \theta = 0.14$ and $v_{sep} = 
0.34$, we obtain $\theta = (50 \pm 11)^\circ$ and $\beta = 0.22 \pm 0.04$. 
Note, however,
that a fully consistent picture of the orientation and kinematics of 
3C\,338 is difficult to obtain: using as a first constraint
the arm length ratio would lead to slightly lower values of $\beta$ and higher
values of $\theta$. 

2) We can assume instead that parsec scale jets are highly relativistic. This 
is supported by previous results on low power radio galaxies (see e.g. 
Giovannini et al. 2001) and on BL Lac sources (Giroletti et al. 2006) which 
according to unified models should be the beamed population of FR I radio 
galaxies. In this case the pattern velocity (proper motion of components E2 
and W2) is not related to the jet bulk velocity and these structures (such as
component E4) are standing shocks as found in other sources (see discussion 
above). Unfortunately we cannot use the core dominance to derive constraints
since the core showed a strong variability and there is clear indication
of a restarted activity. In this last case indeed the core power is not related
to the extended emission which originated in previous active regime.
From the jet/counter-jet ratio and the arm length ratio assuming a jet moving
with Lorenz Factor $\Gamma$ in the range 3 to 10 (see Giovannini et al. 2001) 
the
orientation angle should be in the range 70$^\circ$ -- 80$^\circ$ from the 
line of sight.

\begin{figure*}
\centering
\scalebox{0.80}{\includegraphics[angle=-90]{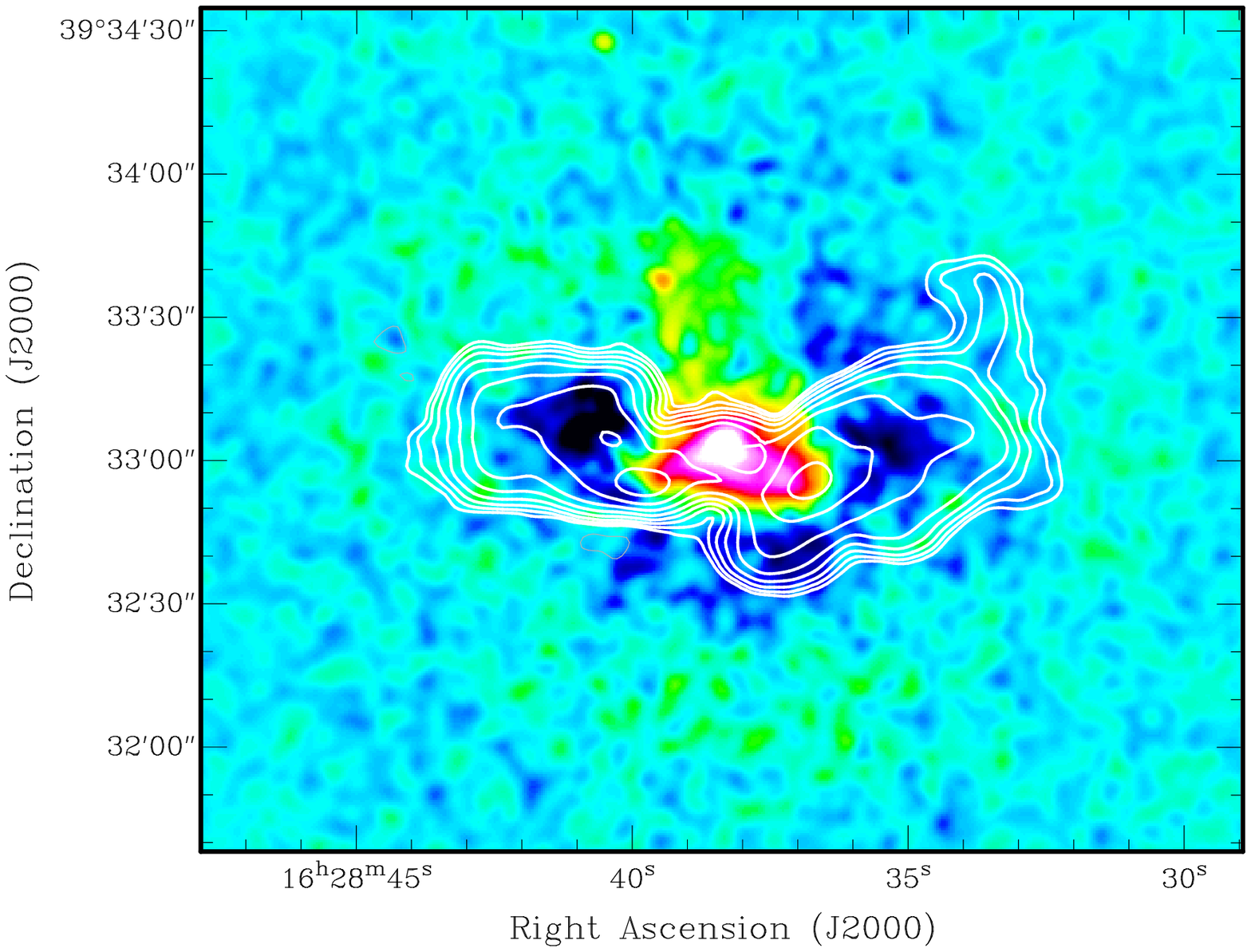}}
\caption{
Unsharped-masked Chandra X-ray image in the 0.5-8.0\,keV band (colours),
formed from the difference of raw images smoothed with Gaussian kernels
of width $\sigma=3$ and 40 pixels, respectively (raw pixel size
$0.492\times0.492$ arcsec$^2$). The resulting unsharped-masked
image has then been smoothed further with a Gaussian kernel of
$\sigma=2$ pixels.
Contours represent the 330 MHz radio map and contour
levels are the same as Figure \ref{330}.
Note the extension to the south, which corresponds to a cavity in the 
X-ray emission that has not been seen before. }
\label{xray_radio}
\end{figure*}

\subsection{Jet Power in 3C\,338}

X-ray studies have shown that the black holes at the centers of galaxies, groups, and clusters are 
capable of inflating cavities or `bubbles' in the surrounding X-ray emitting gas (e.g. Fabian 
\textit{et al}. 2003, 2005 \nocite{Fabian03, Fabian05}; B\^{\i}rzan \textit{et al}. 2004 
\nocite{Birzan04}; Forman \textit{et al}. 2005 \nocite{Forman05}).

Figure \ref{xray_radio} shows an X-ray image overlayed with the contours of the 330 MHz radio map (Figure \ref{330}). The core of the radio emission in 3C\,338, though barely distinguishable at 330  
MHz, is clearly identifyable with the highest X-ray brightness region  
and therefore with the center of the cluster. 
Similar to many recent examples in the literature, the correspondence
between the location of the radio lobes and X-ray cavities (Johnstone et al.
2002) suggests an expansion of the former into the surrounding X-ray
emitting gas.

We estimated the energy required to inflate the large scale bubbles in 3C\,338, based on the results presented by \cite{Allen06}; they studied the properties of the bubbles inflated by the inner lobes,
which are still attached to the jets (see their Fig. 3) and thus likely to still be subjected
to inflation. 
At a projected distance of $\sim 20$ kpc from the core, the electron density is $\sim 0.02$ cm$^{-3}$ and the temperature is $kT \sim 3$ keV. With these values we were able to calculate the thermal pressure, $P$, of the surrounding X-ray emitting gas. 
The energy $E$ required to create the observed bubbles in the X-ray emitting gas (e.g. Allen et al. 2006)
is approximately:

\begin{equation}
E \sim 4PV,
\end{equation}

\noindent
where $V$ is the volume of the cavity. We have modelled the bubbles as ellipsoids with volumes 
$V = 4 \pi r_l r_w ^2 /3$, where $r_1$ is the semi-axis length along the jet direction, and $r_w$ 
is the semi-axis width across it.
Using the 330 MHz image (see Figure \ref{330}) as a guide
to estimate $r_l$ and
$r_w$, we estimate energies require to inflate the large-scale cavities of $E \sim 2.5 \times 10^{59}$ erg and 
$E \sim 3.7 \times 10^{59}$ erg 
for the eastern and western large-scale lobes, respectively.
These values are $\sim 300$ times larger than the one obtained by \cite{Allen06} for the inner bubbles.

For a comparison, assuming equipartition conditions with the ratio between 
electrons and protons k = 1 and a filling factor $\phi$ = 1 we find
in the same volume a minimum energy $\sim$ 3 $\times$ 10$^{58}$ erg and
a magnetic field H$_{eq}$ $\sim$ 15 $\mu$G.

We also estimated the age of the bubbles as (see e.g. B\^{\i}rzan \textit{et al}. 2004 \nocite{Birzan04}) 

\begin{equation}
t_{age} = R/c_s,
\end{equation}

\noindent
where $R$ is the distance of the bubble center from the black hole, and $c_s$ is the adiabatic sound speed of the gas at the bubble radius. Using $R$ estimated from our 330 MHz image ($\sim$ 20 kpc for both bubbles),
we obtain $t_{age} \sim 2.2 \times 10^7$ yr, in good agreement with the
radiative age estimated assuming equipartition conditions ($\sim$ 4 $\times$ 10$^7$ yrs).
Finally, the power involved in `blowing' the bubbles can be estimated from 

\begin{equation}
P_{jet} = E/t_{age}.
\end{equation}

Using our values obtained for $E$ and $t_{age}$ we get $P_{jet} \sim 4.3 \times 10^{44}$ erg s$^{-1}$.

Another estimate can be given using the buoyancy time $t_{buoy}$ instead of $t_{age}$, as the bubbles might have
separated from the AGN. 
It can be calculated from the buoyancy velocity $v_{\rm b}=\sqrt{2gV/SC_{\rm D}}$,
where $g=GM(<R)/R^2$ is the gravitational acceleration at radius R (the distance between
the cluster core and the bubble center), $V$ is the volume,
$S=\pi r_{\rm w}^2$ is the area of the bubble in the rise direction, 
and $C_{\rm D}$ is the drag coefficient, whose value is $\sim 0.75$.
Then, the buoyancy time can be computed from $t_{buoy}=R/v_{\rm b}$.
We find $t_{buoy}=2.4 \times 10^7$ yr for the east bubble and
$t_{buoy}=2.1 \times 10^7$ for the west bubble, values which are very similar to $t_{age}$.
Using $t_{buoy}$ we get $P_{jet} \sim 4.0 \times 10^{43}$ erg s$^{-1}$ and $P_{jet} \sim 4.7 \times 10^{44}$ erg s$^{-1}$
for the eastern and western outer bubbles.
\cite{Allen06} obtained a value of $P_{jet} = (0.741 \pm 0.316) \times 10^{43}$ erg s$^{-1}$ for the inner bubble, which is $\sim 60$ times smaller than the value that we obtain for the large scale bubble. \cite{Allen06} also calculated the accretion power, obtaining $P_{Bondi} = (3.06 ^{+3.65} _{-1.36})\times 10^{43}$ erg s$^{-1}$. The fact that the current value for the accretion power is smaller than the power necessary to inflate the older, large-scale bubble suggests that the accretion rate or accretion mode has varied with time. We also compared our results for $E$ and $P_{jet}$ with those obtained by \cite{Dunn04} for the large scale bubbles and verified that they show good agreement. 
Minor differences are due to slightly different conventions adopted in the two studies.

\begin{deluxetable}{lccc}
\tabletypesize{\scriptsize}
\tablecolumns{4}
\tablewidth{0pt}
\tablecaption{Continuum Spectrum Results\tablenotemark{*}.\label{Peak_Fluxes}}
\tablehead{\colhead{Component}&\colhead{Frequency}&\colhead{Flux}
&\colhead{$\alpha_{8-15}$}\\
\colhead{} &\colhead{(GHz)}&\colhead{(mJ)} &\colhead{}}
\startdata
C...    & 8.4   & 20.98 $\pm$ 1.08 & $-$0.11 $\pm$ 0.01 \\
        & 15.4  & 19.61 $\pm$ 1.09 &               \\
E2...   & 8.4   & 8.07  $\pm$ 0.43  & $-$0.95 $\pm$ 0.04 \\
        & 15.4  & 4.55  $\pm$ 0.34 &               \\
W2...   & 8.4   & 3.97  $\pm$ 0.23  & $-$0.71 $\pm$ 0.07 \\
        & 15.4  & 2.59  $\pm$ 0.24  &                \\
\enddata
\tablenotetext{*}{Results obtained from the 1998 observations. 
The flux densities were measured from matching resolution images.}
\label{spectra}
\end{deluxetable}







\section {Conclusions}

We analysed VLA and VLBA observations of the radio source 3C\,338,
associated with the cD galaxy NGC 6166, the central dominant galaxy of the 
galaxy cluster Abell 2199.
The VLBA observations probe the innermost
region close to the jet with a resolution of the order of 1 pc.
The VLA observations, performed at 0.074, 0.330, and 8.4 GHz, trace the large-scale
emission from the radio lobes out to $40 - 50$ kpc.

The VLBA data reveal the pc-scale jets,
whose kinematics and orientation we did not manage to derive
in an unambiguous way. 
The observations suggest two possible explanations: either the jets are strongly relativistic
and the jets lie within 10\deg$ - $20\deg of the plane of the sky,
or they are only mildly relativistic, and they are pointing 
at an angle of 30\deg$ - $50\deg.

The arcsecond resolution VLA data enable us to investigate the 
large scale structure of the radio source. The morphology of the 
low frequency radio lobes clearly indicates that they are 
associated with the cavities present in the X-ray emission.
The low frequency data also reveal the presence of an extension
to the south which corresponds to an X-ray hole.
We computed the age of the bubbles based on the sound speed of the
surrounding gas, the buoyancy time, and the radiative age:
they are all in fair agreement.
Estimates of the power necessary to blow these cavities suggest
that the accretion power onto the central engine has not been constant
over time.

\begin{acknowledgements}
We thank the referee for constructive comments that improved the 
quality of the paper.
GBT acknowledges support for this work from the
National Aeronautics and Space Administration through Chandra Award
Number GO4-5134A issued by the Chandra X-ray Observatory Center, which
is operated by the Smithsonian Astrophysical Observatory for and on
behalf of the National Aeronautics and Space Administration under
contract NAS8-03060.
GBT is also grateful for hospitality at the CNR in Bologna where a lot
of this work was performed.
Basic research in radio astronomy at the Naval Research Laboratory is
supported by 6.1 base funding.

\end{acknowledgements}



\end{document}